\documentclass{ws-ijmpe}

\begin{document}

\markboth{T. Tarnowsky, R. Scharenberg, B. Srivastava}{Energy and System Size Dependence Study of the Percolation Phase Transition}

\catchline{}{}{}{}{}

\title{ENERGY AND SYSTEM SIZE DEPENDENCE STUDY OF THE PERCOLATION PHASE TRANSITION
}

\author{\footnotesize TERENCE TARNOWSKY, ROLF SCHARENBERG, BRIJESH SRIVASTAVA\\ (for the STAR Collaboration)}

\address{Physics Department, Purdue University, 525 Northwestern Ave.\\
West Lafayette, IN 47906,
USA\\
tjt@physics.purdue.edu}

%
%
\maketitle

\begin{history}
\received{(received date)}
\revised{(revised date)}
\end{history}

\begin{abstract}
Multiparticle production at high energies is described in terms of color strings stretched between the projectile and target. As the string density increases, overlap in the transverse plane leads to cluster formation. At some critical density a macroscopic cluster appears, spanning the entire system. This marks the percolation phase transition. Data from $\sqrt{s_{NN}}$ = 200 GeV p+p, d+Au and Au+Au collisions at RHIC has been analyzed using the STAR detector to obtain the percolation density parameter, $\eta$. 
The particle p$_{T}$ spectrum from 200 GeV p+p data is parameterized using a power law. Values of the fit parameters are used in the d+Au and Au+Au analysis. For 200 GeV Au+Au collisions, the value of $\eta$ is found to lie above the critical percolation threshold, while for other collision systems and energies, it is below the critical value. This supports the idea of string percolation, which at high enough string density is a possible mechanism to explore the hadronic phase transition to a quark-gluon plasma. 

\end{abstract}

\section{Introduction}

It is postulated that in the collision of two nuclei at high-energy, color strings are formed between projectile and target partons. These color strings decay into additional strings via  \begin{math} 
	q-\overline{q}
\end{math}
production, and ultimately hadronize to produce the observed hadrons.\cite{1}

In the collision process, it is possible for partons from different nucleons to begin overlapping in the plane transverse to the direction of travel of the colliding nuclei.  This overlap probability increases with increasing energy and atomic number of the colliding nuclei.\cite{5} It is this overlap that leads to cluster formation in the transverse plane, similar to the disk model of 2-D percolation theory.\cite{5} At a critical cluster density, a cluster will form that spans the entire system. This is referred to as the maximal cluster and marks the onset of the percolation threshold. The quantity $\eta$, the percolation density parameter, can be used to describe overall cluster density. It can be expressed as 
\begin{equation}\eta = \frac{N\pi r_{0}^{2}}{S}\end{equation} 
with N the number of strings, S the total nuclear overlap area, and $\pi r_{0}^{2}$ the transverse area of the discs/strings. At some critical value of $\eta = \eta_{c}$, the percolation threshold is reached. $\eta_{c}$ is referred to as the critical percolation density parameter. In two dimensions, for a uniform string density, and in the continuum limit, $\eta_{c}$ = 1.175.\cite{4} In terms of phenomena that are considered ``localized'' within a percolating system, such as J/$\Psi$ suppression, the critical value of $\eta_{c}$ is predicted to be 1.72.\cite{6}
\\
To calculate the percolation parameter, $\eta$, a parameterization of pp events at 200 GeV is used to compute the $p_{T}$ distribution 
\begin{equation}\frac{dN}{dp_{T}^{2}} = \frac{a}{(p_{0}+p_{T})^{n}}\end{equation}
where a, $p_{0}$, and n are parameters fit to the data. This parameterization can be used for nucleus-nucleus collisions if one takes into account the percolation of strings by\cite{1}
\begin{equation}p_{0} \longrightarrow p_{0}\left( \frac{\left<\frac{nS_{1}}{S_{n}}\right>_{Au-Au}}{\left<\frac{nS_{1}}{S_{n}}\right>_{pp}}\right)^{\frac{1}{4}}\end{equation}

In pp collisions at 200 GeV, the quantity $\left<\frac{nS_{1}}{S_{n}}\right>_{pp} = 1.0 \pm 0.1$, due to the low probability of string overlap in pp collisions. Once the $p_{T}$ distribution for nucleus-nucleus collisions is determined, the multiplicity damping factor can be defined in the thermodynamic (continuum) limit as \cite{5}
\begin{equation}\label{eq4}F(\eta) = \sqrt{\frac{1-e^{-\eta}}{\eta}}\end{equation}
which accounts for the overlapping of discs, with $1-e^{-\eta}$ corresponding to the fractional area covered by discs. The quantity $\left<\frac{nS_{1}}{S_{n}}\right>_{Au-Au}$ is equal to $F(\eta)^{-2}$.

\section{Data Analysis}

The data utilized in this analysis was acquired by the STAR experiment at RHIC. Only data from $\pm$ 1 unit of pseudorapidity with greater than 10 fit points in the main STAR detector, the Time Projection Chamber, was used. The events studied were minimum bias events, with standard STAR centrality cuts as defined by the trigger detectors and overall event multiplicities, which increase with increasing collision centrality. The longitudinal (z) vertex was constrained to within $\pm$ 30 centimeters of the primary event vertex. The distance of closest approach (dca) of primary tracks to the event vertex was less than 3.0 cm. All soft, charged particles with transverse momentum less than 1.2 GeV were considered.

The collision systems studied includes: Au+Au at 200, 62.4, and 19.6 GeV and pp, d+Au, and Cu+Cu at 200 GeV.

\section{Results and Discussion}

The percolation density parameter, $\eta$, has been determined for several colliding systems and energies.\cite{10,11} These results have been compared to a predicted value of the critical percolation density, $\eta_{c}$. If the critical value of $\eta$ is exceeded, it is expected that the percolation threshold has been reached. This would then indicate the formation of a maximal cluster that spans the system under study. All results are shown with statistical errors only, which are smaller than the data points.
\\
\begin{figure}
\centering
\includegraphics[width=4in]{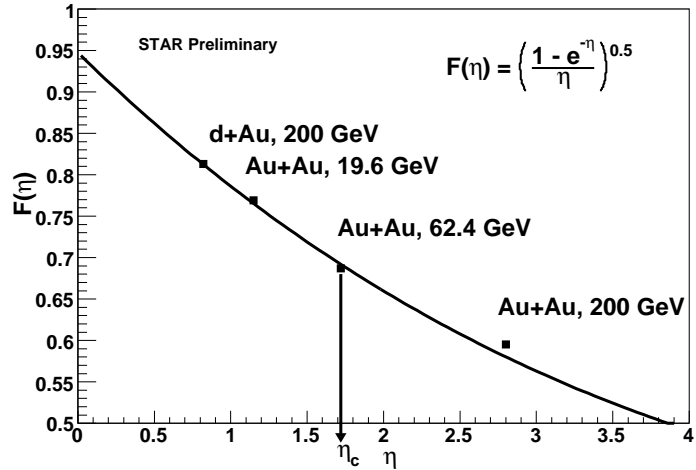}
\caption{\footnotesize{Multiplicity suppression factor, F($\eta$) versus the percolation density parameter, $\eta$. The line is drawn to guide the eye, not as a fit to the points. The estimated localized critical percolation density for 2-D overlapping discs, $\eta_{c}$, is shown. The $\eta$ values are calculated for the most central collisions. The values for 200 GeV d+Au and 19.6 GeV Au+Au lie below $\eta_{c}$, while the result for 200 GeV Au+Au lies above. 62.4 GeV Au+Au falls near the critical value.}}
\label{Fig1}
\end{figure}
Figure 1 is a plot of the quantity F($\eta$) versus the percolation density parameter, ($\eta$), for central collisions. The critical percolation threshold is indicated as that for overlapping strings with a non-uniform density (``local'' value).\cite{6} There is a general increase in $\eta$ with increasing system size and energy. The predicted value of $\eta_{c}$ lies well above that of 19.6 GeV Au+Au and 200 GeV d+Au collisions. The result of 62.4 GeV Au+Au is located near to $\eta_{c}$. Only 200 GeV Au+Au lies well above this predicted percolation threshold. It is expected that the overlap probability for string clusters increases as the system size and energy increase. Therefore, a larger percolation density should be obtained at 200 GeV Au+Au compared to lower energies and other, lighter systems.

One can also consider the percolation density as a function of centrality in Au+Au collisions. The centrality expressed in terms of the number of participating nucleons ($N_{part}$) as found from Monte Carlo Glauber calculations.\cite{7} Larger values of $N_{part}$ correspond to more central collisions. Figure 2 shows $\eta$ as a function of the number of participant nucleons in Au+Au collisions at 200 and 62.4 GeV. It is shown here that for all centralities, except for the most peripheral bins, all 200 GeV Au+Au collisions lie above the critical percolation threshold. The value of $\eta$ increases with increasing collision centrality, an expected indication of additional string overlap in more central collisions. For 62.4 GeV Au+Au, all centralities lie at, or below, the critical percolation density. This is true even for the most central collisions at this energy. Another interesting feature is that the value of $\eta$ appears to achieve a maximum value of approximately 1.7 in central 62.4 GeV Au+Au collisions, whereas in central 200 GeV Au+Au collisions $\eta$ continues to increase. This could potentially indicate a saturation of the cluster overlap area, and hence the percolation parameter in central 62.4 GeV Au+Au collisions.   

\begin{figure}
\centering
\includegraphics[width=4in]{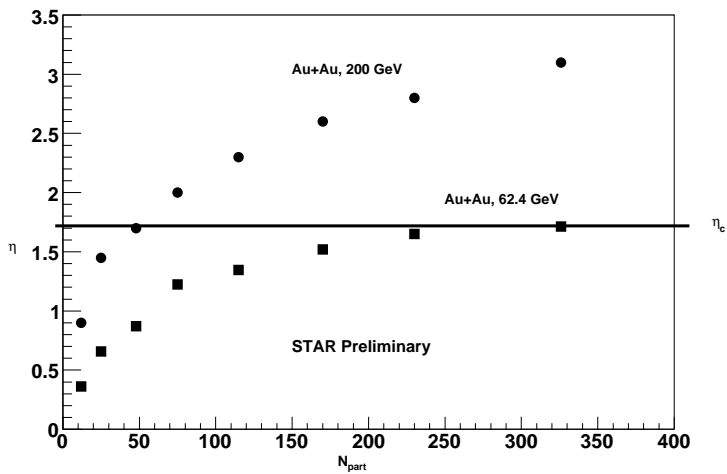}
\caption{\footnotesize{The percolation density parameter, $\eta$, as a function of collision centrality ($N_{part}$) in 62.4 and 200 GeV Au+Au collisions. Larger values of $N_{part}$ correspond to more central collisions. For all but the most peripheral collision centralities, 200 GeV Au+Au exceeds the critical percolation threshold, $\eta_{c}$. In 62.4 GeV Au+Au, all centralities lie at or below $\eta_{c}$.}}
\label{Fig2}
\end{figure}

Finally, Fig. 3 presents the fractional area ($1-e^{-\eta}$) covered by clusters as a function of $\eta$. The data is shown for all centralities of 62.4 and 200 GeV Au+Au collisions, as well as 200 GeV Cu+Cu. The 200 GeV Cu+Cu result spans a region between that of mid-central to central 62.4 GeV Au+Au and mid-peripheral 200 GeV Au+Au. Unlike the Au+Au results, the Cu+Cu data is not corrected for efficiency and detector acceptance. It has been predicted by Satz\cite{4} that the study of Cu+Cu collisions at RHIC energies provides one of the best possibilities to study the onset of deconfinement due to the fact that for heavier nuclei (such as Pb+Pb) at top RHIC energy, all centralities are above the percolation threshold. As shown in Fig. 3, 200 GeV Cu+Cu and 62.4 GeV Au+Au appear to probe the region around the critical percolation threshold.

\begin{figure}
\centering
\includegraphics[width=4in]{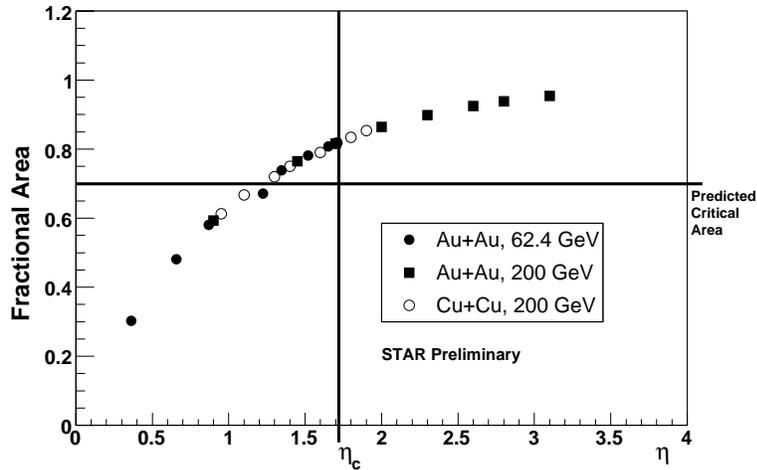}
\caption{\footnotesize{The fractional area covered by clusters as a function of the percolation density parameter, $\eta$. Results for 200 and 62.4 GeV Au+Au, along with 200 GeV Cu+Cu are shown. The new, preliminary Cu+Cu results lay in the regime that spans the range of $\eta$ covered by peripheral 200 GeV Au+Au and central 62.4 GeV Au+Au.}}
\label{Fig3}
\end{figure}

\section{Summary}

In summary, we have presented the preliminary results for the percolation density parameter, $\eta$, at RHIC for several collision systems as a function of collision centrality. Percolation theory describes a phase transition as occurring at a critical percolation density, $\eta_{c}$. These results show that for light systems (d+Au) or heavy systems at low energies (Au+Au at 19.6 GeV), the calculated value for $\eta$ lies well below the predicted critical value. For Au+Au collisions at 200 GeV, $\eta$ lies well above the predicted percolation threshold. However, 62.4 GeV Au+Au data appears to lie at or below this threshold. As a function of centrality in Au+Au collisions at 62.4 and 200 GeV, almost all centralities except the most peripheral for 200 GeV are above the percolation threshold, while for the lower 62.4 GeV events, all centrality bins lie just at, or below, $\eta_{c}$. Results for Cu+Cu collisions at 200 GeV have been compared to 62.4 and 200 GeV Au+Au. The Cu+Cu results span the regime between central 62.4 and peripheral 200 GeV Au+Au. The onset of this collective behavior in traditional percolation theory is indicative of a phase transition. This could indicate the creation of an intermediate phase between a QGP and the hadronic state.\\

\end{document}